\documentclass[]{pasj01}
\usepackage{color}

\begin{document} 
\Received{2015/09/04}
\Accepted{2016/01/04}

\title{Angular Momentum of the N$_2$H$^+$ Cores in the Orion A Cloud}

\author{Ken'ichi \textsc{Tatematsu},\altaffilmark{1,2}
Satoshi \textsc{Ohashi},\altaffilmark{3}
Patricio \textsc{Sanhueza},\altaffilmark{1}
Quang \textsc{Nguyen Luong},\altaffilmark{1,4}
Tomofumi \textsc{Umemoto},\altaffilmark{1,2}
and
Norikazu \textsc{Mizuno}\altaffilmark{1,3}
}
\altaffiltext{1}{National Astronomical Observatory of Japan, 
2-21-1 Osawa, Mitaka, Tokyo 181-8588}
\altaffiltext{2}{Department of Astronomical Science, 
SOKENDAI (The Graduate University for Advanced Studies), 
2-21-1 Osawa, Mitaka, Tokyo 181-8588}
\altaffiltext{3}{Department of Astronomy, The University of Tokyo, Bunkyo-ku, Tokyo 113-0033}
\altaffiltext{4}{East Asian Core Observatories Association (EACOA) Fellow}

\email{k.tatematsu@nao.ac.jp, 
satoshi.ohashi@nao.ac.jp, 
patricio.sanhueza@nao.ac.jp,
quang.nguyen-luong@nao.ac.jp,
umemoto.tomofumi@nao.ac.jp, 
norikazu.mizuno@nao.ac.jp
}


\maketitle

\begin{abstract}
We have analyzed the angular momentum of the molecular cloud cores in the Orion A giant molecular cloud observed in the N$_2$H$^+$ $J = 1\rightarrow0$ line with the Nobeyama 45 m radio telescope. We have measured the velocity gradient using position velocity diagrams passing through core centers, and made sinusoidal fitting against the position angle. 27 out of 34 N$_2$H$^+$ cores allowed us to measure the velocity gradient without serious confusion.  The derived velocity gradient ranges from 0.5 to 7.8 km s$^{-1}$ pc$^{-1}$. 
We marginally found 
that the specific angular momentum $J/M$ (against the core radius $R$) of the Orion N$_2$H$^+$ cores tends to be systematically larger than that of molecular cloud cores in cold dark clouds obtained by Goodman et al., in the $J/M-R$ relation.  The ratio $\beta$ of rotational to gravitational energy is derived to be 
$\beta$ = $10^{-2.3\pm0.7}$, 
and is similar to that obtained for cold dark cloud cores in a consistent definition.  
The large-scale rotation of the $\int$-shaped filament of the Orion A giant molecular cloud does not likely govern the core rotation at smaller scales. 
\end{abstract}

\section{Introduction}
Angular momentum $J$ plays an essential role in the formation of stars (including binary) and planets (e.g., \cite{bod95}).  For molecular clouds and their cores inside, 
it is observationally found 
that the specific angular momentum $J/M$ (angular momentum per unit mass) increases with increasing radius $R$ in the power-law relation of $J/M \propto R^{1.6}$ \citep{goo93,gol85}.  It is suggested that the $J/M-R$ relation is related to the linewidth-size ($\Delta v-R$) relation, one of the empirical relations found by \citet{lar81} \citep{goo93,bod95}.
Because $J/M = I \omega/M = p R v_{rot}$, 
the $J/M-R$ relation can be expressed as $v_{rot} \propto R^{0.6}$, which is very similar to the power-law form of the linewidth-size relation.  
Here, $I$ is the moment of inertia, $v_{rot}$ is the rotation velocity, $\omega$ is the angular velocity, and the parameter $p$ is equal to $\frac{2}{5}$ for a uniform density sphere.   
When we use the observed linewidth $\Delta v$ rather than non-thermal or total linewidth (see \cite{ful92} for definition), the linewidth-size relation is expressed as $\Delta v = A R^a$ with $a \sim$ 0.4$-$0.5, (e.g., \cite{lar81,ful92,bli93}).  Here, $A$ is a constant or coefficient for the power-law relation (corresponding to the intercept in the log-log form, log $\Delta v$-log $R$ relationship). The origin of the linewidth-size relation is still under debate, and many theoretical studies have been carried out.  
For example, \citet{ino12} studied the formation of molecular clouds due to accretion of \textsc{Hi} clouds through magnetohydrodynamic simulations including the effects of radiative cooling/heating, chemical reactions, and thermal conduction, and successfully reproduced the observed linewidth-size relation.  
Molecular clouds and their cores are thought to be near virial equilibrium (e.g., \cite{lar81,mye83}).  The ratio $\beta$ of rotational to gravitational energy is found to be of the order of 0.02 for dark cloud cores \citep{goo93}. \citet{goo93} suggested that the $J/M-R$ relation can be explained for virial equilibrium cores if $\beta$ is constant against the core size.
$\beta$ and the ratio $\alpha$ of the thermal to gravitational energy  is thought to affect the core fragmentation process, which will be related to the frequency of binary and multiple star formation (e.g., \cite{miy84,tsu99}; see also \cite{mat03}).

Galactic molecular clouds can be divided into two categories: giant molecular clouds (GMCs) and cold dark clouds (excluding infrared dark clouds here) (e.g., \cite{shu87,tur88,ber07}).  These two categories of clouds show different ranges of cloud mass, and star-formation modes: GMCs are most likely associated with cluster star formation including massive stars, while cold dark clouds show preferentially isolated low-mass star formation.  Furthermore, molecular cloud cores inside these clouds show different characteristics: Warm-temperature, turbulent cores in GMCs and cold thermal cores in cold dark clouds.
It was suggested that the coefficient $A$ (or intercept in the log-log form) of the linewidth-size relation differs between cores in these two categories. \citet{tat93} have studied molecular cloud cores in the Orion A GMC in CS $J = 1\rightarrow0$, and suggested that this coefficient $A$ (or intercept) is larger in Orion cores compared with that for cores in cold dark clouds.  They argued that a larger coefficient means that Orion cores have higher external pressure and/or stronger magnetic fields.  \citet{cas95b} have also suggested different intercepts (and different slopes) between GMC cores and cold dark cloud cores using the Orion A GMC data of \citet{tat93} and other complementary data.  Originally, \citet{lar81} pointed out that the linewidth is relatively small in the Taurus cold dark cloud, larger in the $\rho$ Ophiuchus complex, and even larger in the Orion A GMC (meaning different intercepts).  \citet{tat99} compared the linewidth-size relation between ``well-defined'' Orion A GMC cores and cold dark cloud cores (see their Figure 7) and concluded that the intercept is clearly different, but the power-law index is not so different.  Further evidence of variation of intercept was obtained toward the Galactic Center, where the intercept is even larger than that for Orion cores \citep{tsu12}.  
\citet{hey09}, \citet{bal11}, and \citet{tra16} have also shown that the coefficient (or intercept) of the linewidth-size relation is highly related to the surface density (higher external pressure and/or stronger magnetic fields lead to higher surface density), and it has deviations among clouds.  
It seems that different coefficients (or intercepts) of the linewidth-size relation in different categories of Galactic clouds are well established.

Given that intercept of the linewidth-size relation differs among Galactic clouds, we wonder whether the $J/M-R$ relation also shows any difference. \citet{tat99} compared the $J/M-R$ relation of the Orion CS cores with that for cold dark cloud cores derived in NH$_3$ by \citet{goo93}, and found that the Orion cores may have slightly larger $J/M$. However, their results were not conclusive because of large data scattering and insufficient linear-scale resolution in pc. In this paper, we re-investigate the difference in the $J/M-R$ relation by using data obtained in N$_2$H$^+$ at higher angular resolution toward the Orion A GMC \citep{tat08}.  N$_2$H$^+$ and NH$_3$ are thought to trace similar volumes of dense gas (\S3.5).

In this work we adopt a distance of 418$\pm$6 pc for the Orion A GMC, based on
the work of \citet{kim08}, as the best estimate rather than 450 pc used in \citet{tat08}. At this distance, 1$\arcmin$ corresponds to 0.122 pc.  We correct the core radius (in pc) and core mass of \citet{tat08} decreasing by 6.5\% and  12.5\%, respectively. 

Table \ref{tab:catalog} lists the N$_2$H$^+$ cores in the Orion A GMC from \citet{tat08}.
The kinetic temperature $T_k$ is calculated from the rotation temperature from NH$_3$ observations of \citet{wil99} and by using the conversion given in \citet{dan88}.  In general, N$_2$H$^+$ core positions do not match NH$_3$ observed positions, so we take the value of the nearest NH$_3$ position.   The total linewidth $\Delta v_{TOT}$, including both thermal and non-thermal contribution,  is calculated by using $T_k$ and by correcting for the difference of the mean molecular mass (2.33 u) and the mass of the observed molecule (N$_2$H$^+$, 29 u) \citep{ful92}. 
The core mass $M$ is taken from \citet{tat08}, which was obtained by assuming local thermodynamic equilibrium (LTE) and corrected for the distance revision.  
The virial mass $M_{vir}$ is defined as 210 $R$ (pc) $\Delta v_{TOT}^2$ for a uniform density sphere \citep{mac88}.  The virial parameter $\alpha_{vir}$ is defined as the ratio of the virial mass $M_{vir}$ to the core mass $M$, and its average for Orion N$_2$H$^+$ cores is 0.79$\pm$0.61 (log $\alpha_{vir}$ = -0.21$\pm$0.31).  They are not far deviated from virial equilibrium.
The blank in the $T_k$ column means that we do not have a good nearest NH$_3$ counterpart. The blank in the $M$ column (and then $\Delta v_{TOT}$, $M_{vir}$, and $\alpha_{vir}$ columns)  means either no $T_k$ value or failure in the N$_2$H$^+$ hyperfine fitting to restrict the optical depth.

\newpage

\section{Data and Method}
We use the data of N$_2$H$^+$ $J = 1\rightarrow0$ cores \citep{tat08} obtained with the 45 m radio telescope of the Nobeyama Radio Observatory (NRO). \footnote{Nobeyama Radio Observatory is a branch of the National Astronomical Observatory of Japan, National Institutes of Natural Sciences.}
The half-power beam width of the telescope was 17$\farcs$8$\pm$0$\farcs$4, and the spacing grid employed in mapping observations was 20$\farcs$55.  The spectral resolution was 37.8 kHz ($\sim$ 0.12 km s$^{-1}$).  
The core radius $R$ is $\gtrsim$ 25$\arcsec$ or 0.05 pc, which is thought to imply the detection limit in the observations.  
Details of the observations and examples of N$_2$H$^+$ spectra can be found in \citet{tat08}.  It is known that N$_2$H$^+$ traces the quiescent gas, and is less affected by depletion \citep{ber01,ber02} or by star-formation activities such as molecular outflows (e.g., \cite{wom93}).  Therefore, this molecular line is one of the best molecular lines to be used for this study.  On the other hand, in warm gas (temperature $>$ 25 K), N$_2$H$^+$ will be destroyed by evaporated CO \citep{lee04,tat14}.

Cores in the Orion A GMC are much more crowded compared with cores in dark clouds.  If we use the core number surface density of H$^{13}$CO$^+$ cores \citep{ike07,oni02}, the Orion A GMC and the Taurus molecular cloud, which is a crowded example of cold dark clouds, have 5 and 0.2 H$^{13}$CO$^+$ cores per pc$^2$, respectively, with a difference of the order of 25.  Therefore, core identification and measurement of the velocity gradient are more difficult in the Orion A GMC.

We measured the velocity gradient as follows. 

\begin{figure}
\begin{center}
  \includegraphics[width=10cm]{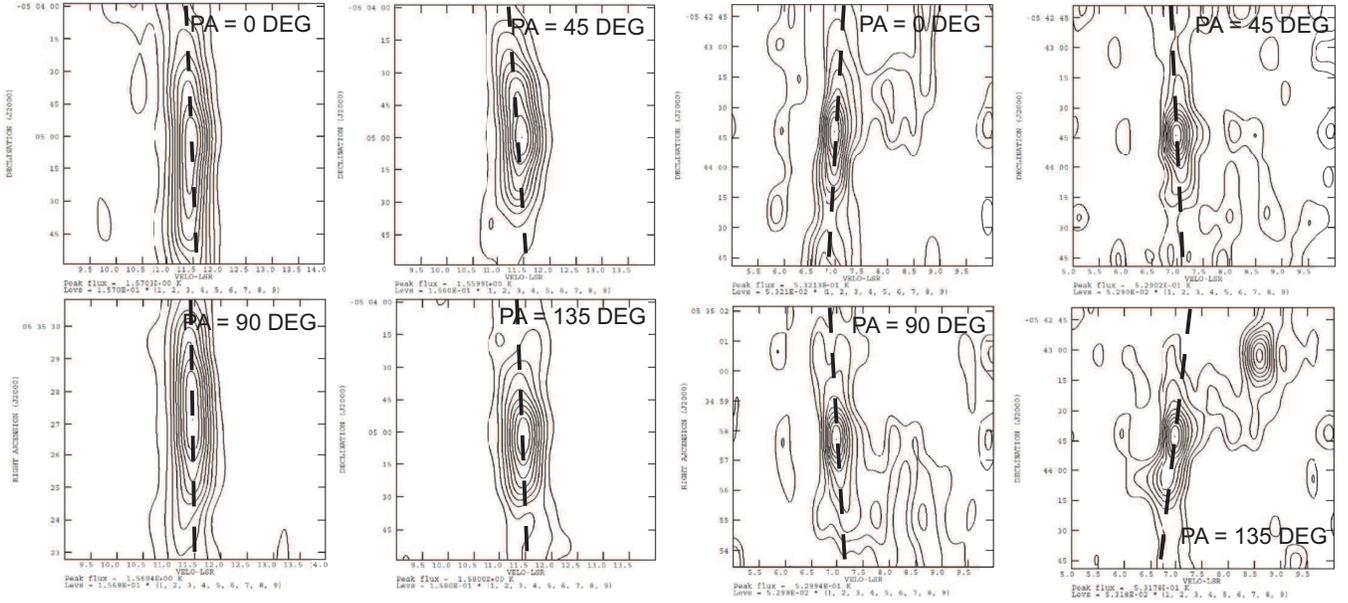} 
 \end{center}
\caption{Position-velocity diagram for N$_2$H$^+$ core 7.  The abscissa is the LSR velocity (km s$^{-1}$) corresponding to N$_2$H$^+$ $J = 1\rightarrow0$,
$F_1$, $F$ = 0, 1$\rightarrow$1, 2.  
The thick dashed line shows the velocity gradient measured through visual inspection.
}\label{fig:pv007n}
\end{figure}

\begin{figure}
 \begin{center}
  \includegraphics[width=10cm]{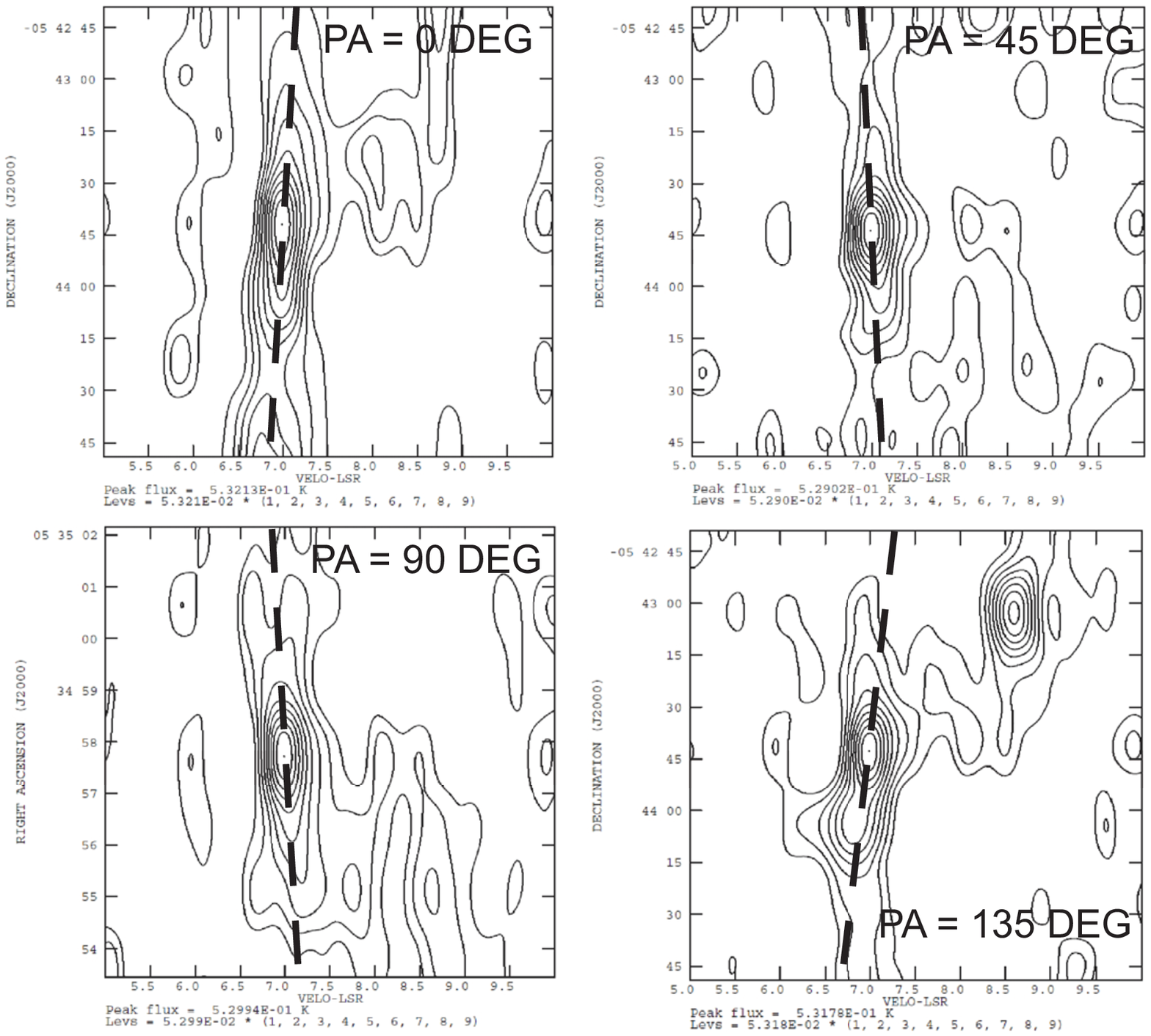} 
 \end{center}
\caption{Same as Figure 1 but for N$_2$H$^+$ core 26.}\label{fig:pv026n}
\end{figure}

 (a) We draw four position-velocity (P-V) diagrams (position angles PA = 0, 45, 90, and 135 degrees) passing through the core center.\footnote{
PV diagrams will be available through the website http://alma.mtk.nao.ac.jp/~kt/kt-e.html}  
We use the isolated hyperfine component N$_2$H$^+$ $J = 1\rightarrow0$,
$F_1$, $F$ = 0, 1$\rightarrow$1, 2, which is the line less affected by neighboring satellites \citep{cas95a}, to avoid confusion.  
Because the data was taken only on a regular grid (rather than on-the-fly mapping observations), the number of actually observed positions along the strip line depends on PA. Interpolation would introduce method-dependent uncertainties that we want to avoid.
The interval of actually observed positions is 20$\arcsec$ for PV diagrams at PA = 0$\degree$ and 90$\degree$, and 28$\farcs$28 (= 20$\sqrt{2}\arcsec$) at PA =  45$\degree$ and 135$\degree$.
The average HWHM radius of N$_2$H$^+$ cores in Orion A GMC by \citet{tat08} is 39$\arcsec$.  If we draw PV diagrams at PA = 26$\fdg$57 (= $\arctan$(0.5)), 63$\fdg$43, 116$\fdg$57, and 153$\fdg$44, the interval of actually observed positions is 44$\farcs$72 (= 20$\sqrt{5}\arcsec$), which is larger than the average core radius.
Then, only PV diagrams at PA = 0$\degree$, 45$\degree$, 90$\degree$, and 135$\degree$ contain a sufficient number of observed positions along strips.
In summary, we prefer to use the strip line passing through actually observed grid positions, to reduce uncertainties caused by interpolation. Figures \ref{fig:pv007n} and \ref{fig:pv026n} show examples of P-V diagrams.

(b) Next, we measure the velocity gradient mainly using the half (50\%) intensity peak
contour through careful visual inspection, assuming solid body rotation.

(b1) We check whether the half intensity peak
contour is confused by another core or not.
If it is not confused, we determine the velocity gradient line by using the minimum and maximum velocities of the half intensity peak
contour.  

(b2) If the half intensity peak contour is symmetrical with respect to the velocity gradient line, we adopt it. 

(b3) If the half intensity contour is not smooth (i.e., irregular, distorted, or wavy shape) near the minimum or maximum velocity point, we try to determine the velocity gradient line so that the half intensity contour looks symmetrical to the line.  

(b4) We also check the 
60\%, 70\%, and 80\% intensity peak contours.  If their minimum and maximum velocities are fitted by another velocity gradient line, and if 60\%, 70\%, and 80\% contours are more symmetrical with respect to this velocity gradient line, we adopt it.  The velocity gradient line does not necessarily pass through the core center, but we allow this if the candidate contour looks like an oval rather than a dogleg or very irregular shape as a whole. 

We have measured the velocity gradient in 27 cores (out of 34 cores) without serious confusion with other cores/emission features spatially and in velocity. 
 
\begin{figure}
 \begin{center}
  \includegraphics[width=10cm]{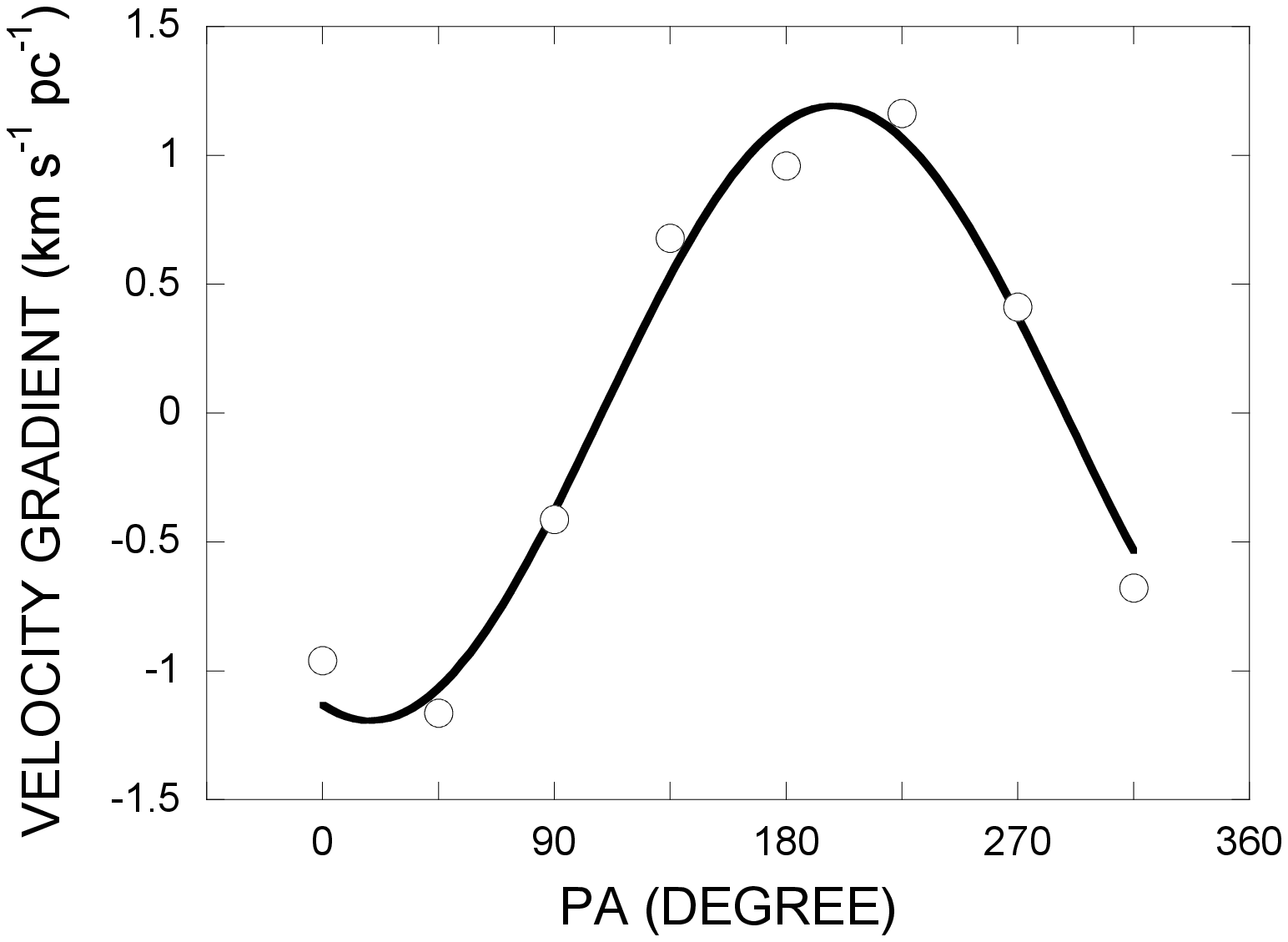} 
 \end{center}
\caption{ Sinusoidal fit to the velocity gradient fit against PA for N$_2$H$^+$ core 7.}\label{fig:vgrad007}
\end{figure}

\begin{figure}
 \begin{center}
  \includegraphics[width=10cm]{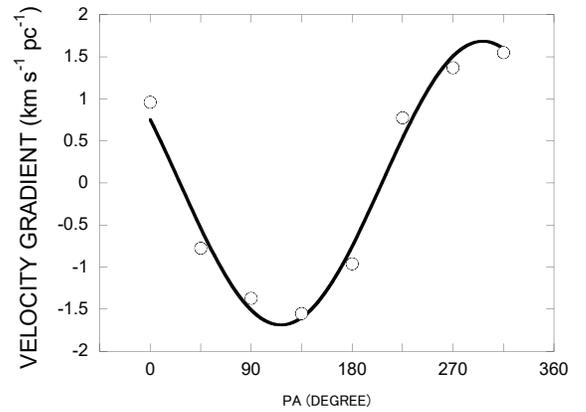} 
 \end{center}
\caption{ Sinusoidal fit to the velocity gradient fit against PA for N$_2$H$^+$ core 26.}\label{fig:vgrad026}
\end{figure}

(c) We fit a sinusoidal curve to the diagram of velocity gradient against PA through non-linear least-squares fitting. 
The amplitude of the sinusoidal function provides the intensity of the velocity gradient, while the angle at which the maximum is found, describes the position angle of the velocity gradient. 
Figures \ref{fig:vgrad007} and \ref{fig:vgrad026} show examples of the sinusoidal fitting.  The velocity gradient on the position-velocity diagram for each PA (0, 45, 90, and 135 degrees) is plotted twice at PA and PA + 180$\degree$ with opposite signs for gradients. 

Table \ref{tab:table2} lists the result of the measurements.
The N$_2$H$^+$ core number is taken from \citet{tat08}.  A velocity gradient of ``0'' means that we cannot measure the gradient  at each PV diagram because it is too small.  
$\beta$ is the ratio  of rotational to gravitational energy (see \S3.2). 
The blank in the V Grad columns means that the velocity gradient is not measurable without confusion.
The blank in the $\beta$ column in Table \ref{tab:table2} means that the core mass was not accurately estimated in \citet{tat08}.  
The average physical parameters of velocity-gradient measurable cores are 
$T_{ex}$ = 9.5$\pm$4.3 K, 
$\Delta v$ = 0.82$\pm$0.44 km s$^{-1}$, 
$R$ = 0.082$\pm$0.026 pc, and
$M$ = 42$\pm$28 $M_{\odot}$.  The velocity gradient derived for 27 cores ranges from 0.5 to 7.8 km s$^{-1}$ pc$^{-1}$, and its average is 2.4$\pm$1.6 km s$^{-1}$ pc$^{-1}$.

\section{Results and Discussion}
\subsection{Specific Angular Momentum against Radius}
\begin{figure}
 \begin{center}
  \includegraphics[width=10cm]{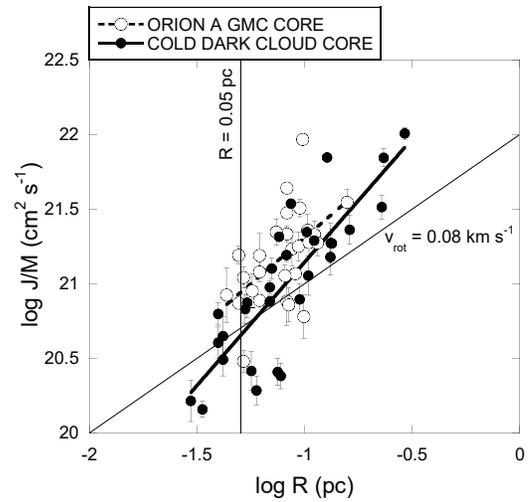} 
 \end{center}
\caption{$J/M-R$ diagram for 27 Orion cores (open circles) and cold dark cloud cores (filled circles).  
The error bar represents uncertainties in measurements of the velocity gradient.
The dashed and solid straight lines are computed using a linear least-squares program for Orion cores and cold dark cloud cores, respectively.  
Thin straight lines delineate approximate detection limit for Orion cores: $R$ = 0.05 pc, and $v_{rot}$ = 0.08 km s$^{-1}$
}\label{fig:JM-R}
\end{figure}

We define the specific angular momentum as $J/M = I \omega/M = \frac{2}{5} R v_{rot}$ for a uniform density sphere.  
Because the moment of inertia $I$ is proportional to mass $M$, $J/M$ can be derived without using the mass estimate.
In figure \ref{fig:JM-R}, we plot the specific angular momentum against core radius.  The core radius $R$ of Orion cores is measured as $R$ = $\sqrt{S/\pi}$, where $S$ is the area within the half intensity contour, and then corrected for the telescope beam size \citep{tat08}.  The data of cold dark cloud cores are taken from \citet{goo93}.   To do an appropriate comparison, we need to use common definitions for physical parameters.  We therefore convert the geometrical mean of two-dimensional FWHM diameter in \citet{goo93} to the HWHM radius (FWHM/2) $R$ in our study.  \citet{goo93} also corrected the radius for the telescope beam size (deconvolved size).  If the velocity gradient $v_{rot}/R$ is constant against $R$, $J/M$ increases as $R^2$ from the definition. Therefore, we divide $J/M$ listed in \citet{goo93} by a factor of four for consistency. The ratio $\beta$ of rotational to gravitational energy is calculated consistently. Table \ref{tab:table2} lists the results. 

Figure \ref{fig:JM-R} shows that Orion cores are mostly located above the least-squares fit of cold dark cloud cores.  The results of linear least-squares fitting are:
\\
\\
log $J/M$ (cm$^2$ s$^{-1}$) = 21.31$\pm$0.07    + 1.24$\pm$0.38 log ($R$/0.1 pc)    
\hspace{50mm}
 (1)
\\
\\
for Orion cores, 
\\
\\
log $J/M$ (cm$^2$ s$^{-1}$) = 21.15$\pm$0.05    + 1.65$\pm$0.20 log ($R$/0.1 pc)            \hspace{50mm}
(2)
\\
\\
for cold dark cloud cores. The number of Orion N$_2$H$^+$ cores located below the least-squares fitting for cold dark cloud cores is relatively small.  
\citet{tat99} showed very similar results for Orion cores using the CS $J$ = 1-0 mapping data of \citet{tat93}.  
The power-law index is less certain in Orion cores compared with cold dark cloud cores, because of a narrower core radius range.  If we assume a power-law index of 1.65 obtained for cold dark cloud cores also for Orion cores, we obtain

log $J/M$ (cm$^2$ s$^{-1}$) = 21.35$\pm$0.05    + 1.65 log ($R$/0.1 pc)    
\hspace{50mm}
 (3)
\\
\\
for Orion cores.
To confirm the results of these fits, we now only use cores with $R \geq$ 0.05 pc and $v_{rot} \geq $ 0.08 km s$^{-1}$, which are the approximate detection limits in the velocity gradient measurement in Orion, for both Orion cores and cold dark cloud cores.
The beam-deconvolved radius $R$ = 0.05 pc corresponds to 24$\farcs$6, which is 1.4 times the telescope beamsize, and 1.2 times the grid spacing of the observations.  We measure the velocity gradient by using the minimum and maximum velocities on both sides with respect to the core center. The difference between the minimum and maximum velocities is 2 $v_{rot}$, which corresponds to a length of twice the half intensity peak radius. We can measure the velocity difference 2 $v_{rot}$ at half intensity peak down to 2.5$-$3 times the instrumental resolution ($\sim$ 0.12 km s$^{-1}$), and the detection limit is approximately
$v_{rot}$ = 0.08 km s$^{-1}$. 
Figure \ref{fig:JM-R-restrict} shows the $J/M-R$ relation.  The results of linear least-squares fitting are:
\\
\\
log $J/M$ (cm$^2$ s$^{-1}$) = 21.38$\pm$0.06    + 1.23$\pm$0.42 log ($R$/0.1 pc)    
\hspace{50mm}
 (4)
\\
\\
for Orion cores, 
\\
\\
log $J/M$ (cm$^2$ s$^{-1}$) = 21.24$\pm$0.05    + 1.30$\pm$0.23 log ($R$/0.1 pc)            \hspace{50mm}
(5)
\\
\\
for cold dark cloud cores. 
It seems that Orion cores have systematically larger $J/M$ than cold dark cloud cores, although the difference is marginal.  
This trend is similar to what we see in the linewidth-size relation \citep{tat93}. 
Our result may imply that non-thermal motions (turbulence) are related to the origin of angular momentum.  Orion cores and cold dark cloud cores have a similar slope (1.2$-$1.3) in Figure \ref{fig:JM-R-restrict}, but their values are shallower than the value of 1.6 found by \citet{goo93} and \citet{gol85}. This could be the result of a narrow core radius range after the restriction.
In general, least-squares fitting tends to provide a shallower slope, if data scattering is large.  Consider a case where $x$ and $y$ are positively correlated.  We fit $x$-$y$ data to a formula $y = a x + b$, and then fit again the data with the expression $x = c y + b$.  We will obtain $c \sim 1/a$ if $x$ and $y$ are well correlated (the correlation coefficient is close to unity). If data is very scattered (the correlation coefficient is small), $a$ will be smaller than $1/c$.  This is because least-square fitting minimizes ``vertical distances'' between the observed points and the fitted points.  By narrowing the core radius $R$ range, scattering becomes larger.  This explains why Orion cores show a shallower slope in Figure \ref{fig:JM-R}, and why the slope of cold dark cloud cores becomes smaller in Figure \ref{fig:JM-R-restrict}.
If we fix the slope to 1.65 for both the Orion and cold dark cloud cores above the Orion core detection limits (Figure \ref{fig:JM-R-restrict-fix}), we obtain
\\
\\
log $J/M$ (cm$^2$ s$^{-1}$) = 21.41$\pm$0.05    + 1.65 log ($R$/0.1 pc)    
\hspace{50mm}
(6)
\\
\\
for Orion cores, 
\\
\\
log $J/M$ (cm$^2$ s$^{-1}$) = 21.23$\pm$0.05    + 1.65 log ($R$/0.1 pc)            \hspace{50mm}
(7)
\\
\\
for cold dark cloud cores.

\begin{figure}
 \begin{center}
  \includegraphics[width=10cm]{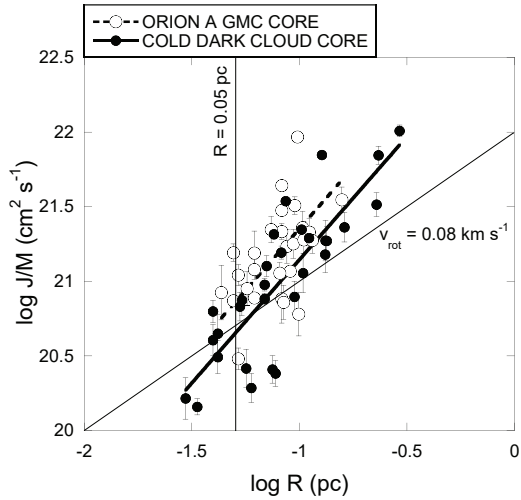} 
 \end{center}
\caption{
Same as Figure \ref{fig:JM-R} but the best-fit line for Orion cores is obtained by fixing a slope (power-law index) of 1.65 obtained for cold dark cloud cores.}\label{fig:JM-R-fix}
\end{figure}

\begin{figure}
 \begin{center}
  \includegraphics[width=10cm]{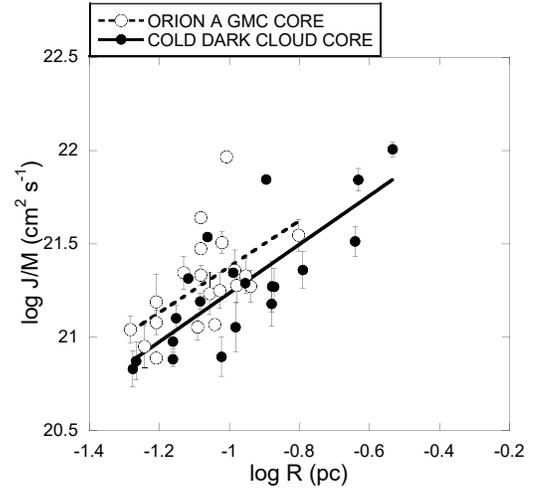} 
 \end{center}
\caption{
Same as Figure \ref{fig:JM-R} but we use only cores with $R \geq$ 0.05 pc and $v_{rot} \geq$ 0.08 km s$^{-1}$.
}\label{fig:JM-R-restrict}
\end{figure}

\begin{figure}
 \begin{center}
  \includegraphics[width=10cm]{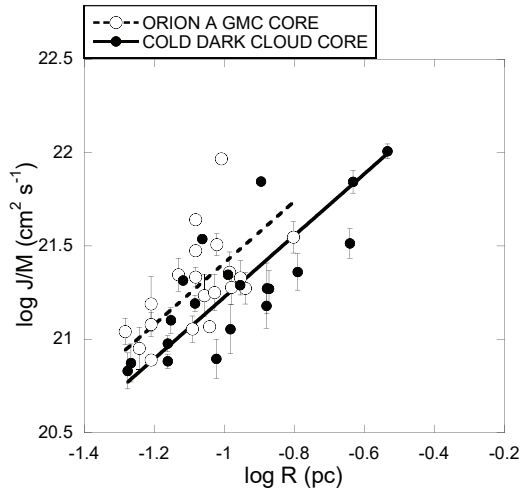} 
 \end{center}
\caption{
Same as Figure \ref{fig:JM-R-restrict} but least-squares fitting is made by fixing the slope to 1.65.}\label{fig:JM-R-restrict-fix}
\end{figure}

\subsection{Ratio of Rotational to Gravitational Energy}
\begin{figure}
 \begin{center}
  \includegraphics[width=10cm]{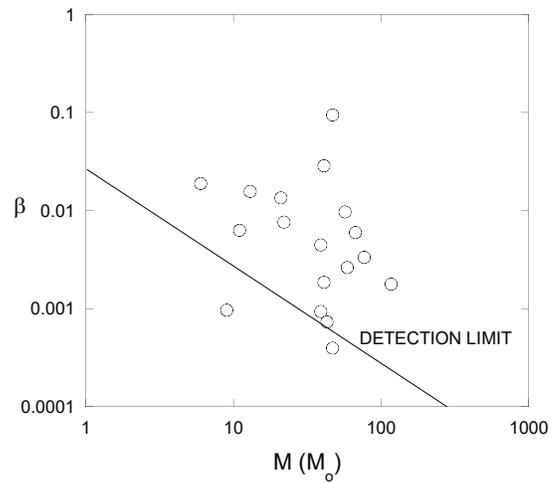} 
 \end{center}
\caption{The ratio $\beta$ of rotational to gravitational energy is plotted against the core mass $M$.  The thin straight line represent the approximate detection limit calculated by using  $R$ = 0.05 pc and $v_{rot} =$ 0.08 km s$^{-1}$.}\label{fig:beta-M}
\end{figure}

The ratio $\beta$ of rotational to gravitational energy is defined as $\beta = \displaystyle \frac{(1/2)I\omega^2}{qGM^2/R} = \frac{1}{2} \frac{p}{q} \displaystyle \frac{\omega^2 R^3}{GM}$ (for a uniform density sphere, $q$ = $\frac{3}{5}$ and $p$ = $\frac{2}{5}$) \citep{goo93}. 
We derived the logarithm of $\beta$ to be $-2.3\pm0.7$ for Orion N$_2$H$^+$ cores (without the restriction above). This is smaller than the value of the logarithm of $\beta$ of $-1.9\pm0.7$ for cold dark cloud cores.  Note that $\beta$ for cold dark cloud cores differs from that in \citet{goo93} due to different definitions of $R$.  If we take the virial mass $M_{vir}$ instead of the core mass, we obtain the logarithm of $\beta$ to be $-2.0\pm0.6$ for Orion N$_2$H$^+$ cores.  
Taking into account a factor of two uncertainty in the absolute estimate of the core mass, we conclude that $\beta$ is similar between Orion and cold dark cloud cores.
We wonder how $\beta$ depends on the core mass for Orion cores.
Figure \ref{fig:beta-M} plots  $\beta$ against the core mass $M$.  It seems that $\beta$ decreases with increasing core mass. It may mean that rotation is more important in low-mass cores.  However, it is likely that this result is affected by the detection limit, because $\beta$ is proportional to $M^{-1}$ and also because cores have core radii and specific angular momenta close to the detection limits.  The thin line in Figure \ref{fig:beta-M} represents the detection limit by using $R$ = 0.05 pc and $v_{rot} =$ 0.08 km s$^{-1}$.

\subsection{Properties of GMC Cores}
GMCs are likely to have deeper gravitational potentials due to their larger masses compared with cold dark clouds (e.g., \cite{tur88}). GMCs have larger pressure than cold dark clouds (e.g., \cite{mye78,tur88}). This may mean that GMCs have higher ratios of molecular to atomic gas \citep{bli06}. Higher external pressure for cores will lead to larger non-thermal motions (turbulence or MHD wave) in cores \citep{chi87}.  It seems that molecular clouds are only slightly supercritical, in various cloud scales from clouds to cores \citep{cru12}, and GMC cores will tend to have stronger magnetic fields.  Therefore, GMC cores will have larger intercepts in the power-law relation of the linewidth-size relation \citep{tat93}. If we assume that turbulence or magnetic fields play an important role in the origin of the specific angular momentum, this might explain why GMC cores tend to have larger specific angular momentum $J/M$.  

Another possibility is that the large overall specific angular momentum in the Orion A GMC \citep{ima11a}, compared with other GMCs, results in a higher $J/M$.  If the global specific angular momentum is related to the local specific angular momentum on scales of molecular cloud cores, cores in the Orion A GMC might have larger specific angular momentum than those in other GMCs.  It is suggested the Orion A GMC has a large angular momentum because it accompanies the richest OB associations \citep{bli90}. The same mechanism may work for Orion cores.

\subsection{Comparison with Global Structure, Global Velocity Gradient, Core Elongation, and Molecular Outflow}
\citet{kut77} measured the velocity gradient along the overall Orion A GMC for the first time, deriving a value of 0.135 km s$^{-1}$ pc$^{-1}$. The velocity gradient was observed along the elongation of the filamentary GMC, and they pointed out that the implied rotation is in the direction opposite to the Galactic rotation.  \citet{bli90} found that the typical angular momentum of Galactic GMCs is less than half that of the Orion A GMC.   The origin of the GMC velocity gradient is discussed in detail by \citet{bli93}, \citet{ima11a}, and \citet{ima11b}, including possibilities of the gradient inherited from the galaxy, that inherited from the parent interstellar medium (\textsc{Hi} clouds), the sweeping action of the stellar association, the influence of magnetic fields including magnetic breaking (e.g., \cite{fie78}), and so on.   
\citet{tat93} found that the $\int$-shaped filament of the Orion A GMC shows a velocity gradient not only along the filament but also across the filament.  The latter observed in the $^{13}$CO $J = 1\rightarrow0$ emission is $\sim$ 2$-$4 km s$^{-1}$ pc$^{-1}$, which is 20$-$40 times larger than the gradient along the filament. We wonder how the velocity gradients of molecular cloud cores inside are.   

\begin{figure*}
  \begin{center}
    \FigureFile(150mm,150mm){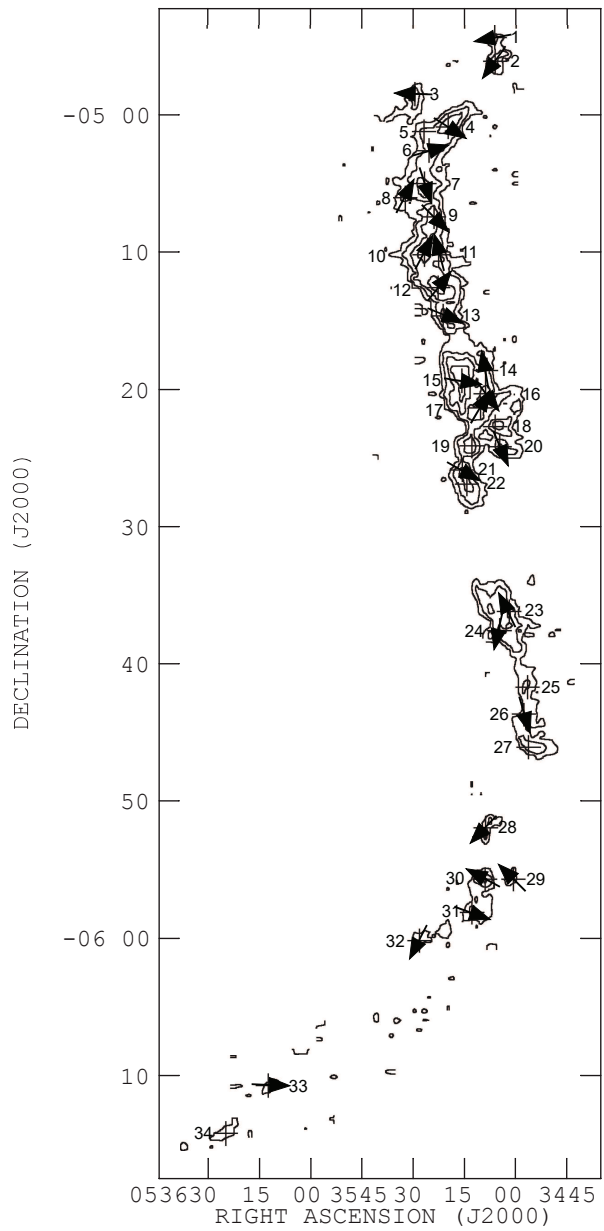}
  \end{center}
  \caption{The orientation of the velocity gradient is plotted on the N$_2$H$^+$ taken from \citet{tat08}. The contour map represents the velocity-integrated intensity 
of the N$_2$H$^+$ 
$J = 1\rightarrow0$ 
$F_1$ = 2$\rightarrow$1 hyperfine group.  The contour levels 
are 0.749 K km s$^{-1} \times$ (1, 2, 4, 8).
The intensity maxima of the N$_2$H$^+$ cores and the orientations of the velocity gradient are shown as
pluses associated with core numbers and arrows, respectively.}\label{fig:n2hp-map}
\end{figure*}

\begin{figure}
 \begin{center}
  \includegraphics[width=10cm]{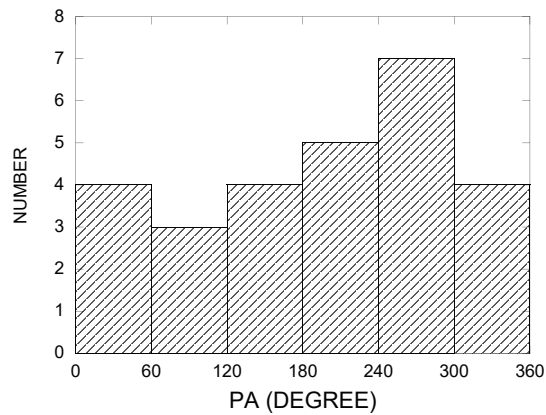} 
 \end{center}
\caption{ Histogram of the 
orientation (position angle) of the velocity gradient 
for the Orion cores.}\label{fig:PA}
\end{figure}

We investigate how the velocity gradient of cores is distributed in the Orion A GMC. 
Figure \ref{fig:n2hp-map} shows 
the orientation of the velocity gradient on the N$_2$H$^+$ map.
In figure \ref{fig:PA}, we plot the histogram of the position angle of the velocity gradient. We define PA = 0$\degree$ if the core has a velocity gradient from south (negative velocity) to north (positive velocity).  The gradient across the filament is a twisting motion with respect to Orion KL \citep{tat93}.
In the north of Orion KL, the eastern side of the filament has a more redshifted velocity, while in the south of Orion KL, the eastern side of the filament has a more blueshifted velocity (see their figure 3).
In our PA definition, PA of the velocity gradient is about 90 degrees in the north of Orion KL and about 270$\degree$ in the south of Orion KL.  N$_2$H$^+$ cores 1-18 are located in the north of Orion KL and N$_2$H$^+$ cores 18-34 are located in the south of Orion KL \citep{tat08}.  There is no clear relation between the filament rotation across the filament and the core velocity gradients.  We conclude that the filament rotation across the filament does not likely govern the core rotation. The global velocity gradient direction along the GMC elongation for the Orion A GMC, which is much smaller than the gradient across the filament, corresponds to PA $\sim 330\degree$ in our definition.  The velocity gradients of cores are not significantly related to the global gradient.     It seems that the origin of the specific angular momentum is not a simple top-down collapse from larger structures (cf. \cite{ima11a} for \textsc{Hi} cloud-GMC comparison).

\begin{figure}
 \begin{center}
  \includegraphics[width=10cm]{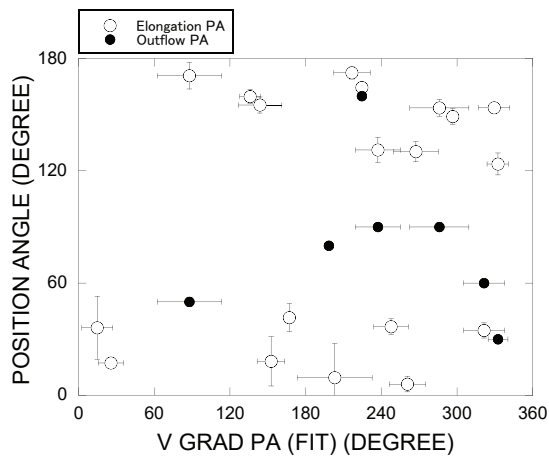} 
 \end{center}
\caption{ Orientations of the elongation and the outflow lobe are plotted against the orientation of the velocity gradient.}\label{fig:PAs}
\end{figure}

\begin{figure}
 \begin{center}
  \includegraphics[width=10cm]{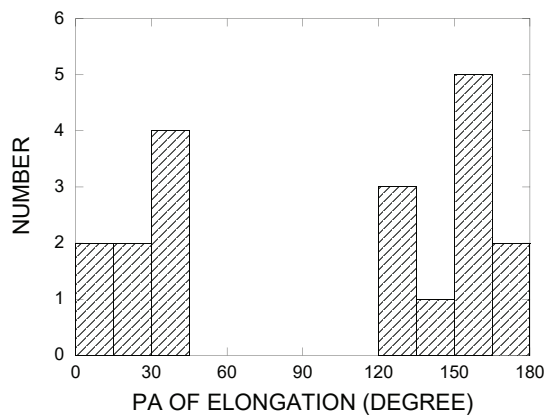} 
 \end{center}
\caption{ Histogram of the core elongation for the Orion cores.}\label{fig:PA-elongation-hist}
\end{figure}

\begin{figure}
 \begin{center}
  \includegraphics[width=10cm]{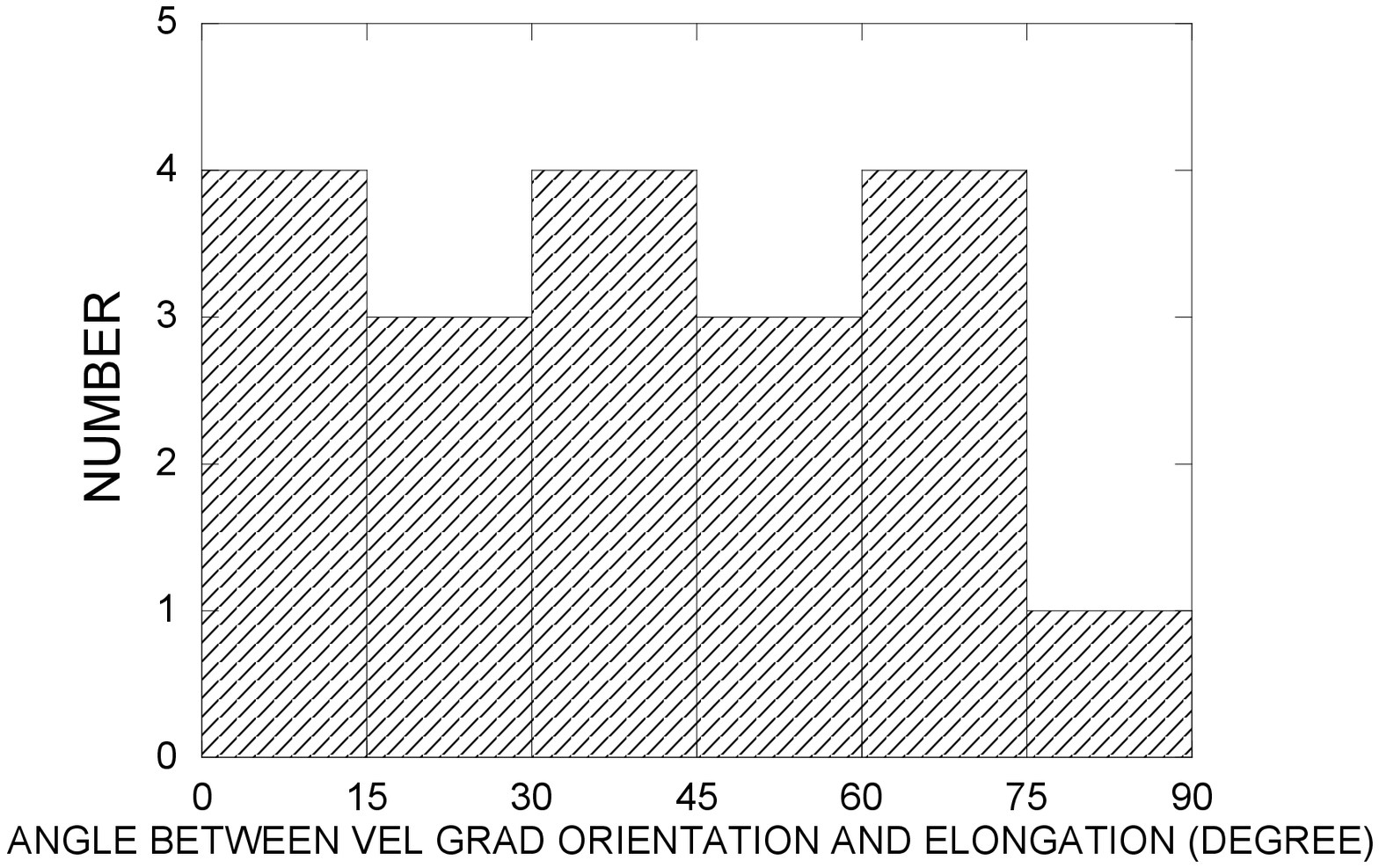} 
 \end{center}
\caption{ Histogram of the angle between the orientation of the velocity gradient and core elongation for the Orion cores.}\label{fig:PA-vgrad-elongation}
\end{figure}

We wonder whether the velocity gradient is related to the core elongation.  If rotation is dominant in core support, these will be positively correlated.
We plot the orientation of the elongation against the orientation of the velocity gradient in figure \ref{fig:PAs}.  
The PA of the core elongation is measured by using the half-intensity contour on the four position-velocity diagrams passing through the core center, and fit the sinusoidal curve on the core size-PA diagram (Table \ref{tab:table3}).  We failed to measure the core elongation in eight cores (out of 27), because they are almost circular or far from elliptical (irregular). In these cases, elongation is blank in the table.  Core elongation has a deficit for PA = 60$\degree$ to 120$\degree$.
Figure \ref{fig:PA-elongation-hist} shows the histogram of the PA of the core elongation.
This means that cores having an elongation perpendicular to the global filament are rare.
In figure \ref{fig:PA-vgrad-elongation}, we plot the histogram of the angle between the velocity gradients and elongations of the cores.  There is a decrease of cores with the velocity gradient perpendicular to the core elongation.  However, it could be due to selection effects, since it is harder to measure the gradient across the elongation due to limited resolution. Except for this depression, there is no clear tendency between the velocity gradient and the core elongation.   Rotationally supported oblate cores will show velocity gradients parallel to the core elongation, but we do not see such a peak. This is naturally understood, because $\beta$ is small.  

Next, we discuss the relationship among the orientation of the velocity gradient, that of the molecular outflow, and that of magnetic fields. \citet{mat04} have studied the relationship between the magnetic field direction and the resulting orientation of outflows/jets theoretically.  They concluded that outflows tend to be aligned with magnetic fields of the parent cloud.    Molecular outflows in the OMC-2/3 region were studied by \citet{aso00}, \citet{yu00}, \citet{shi08}, and \citet{tak08}. In addition, we include a molecular outflow associated with Orion-S observed by \citet{sch90}.  We list molecular outflows (driving source names from \cite{chi97} and \cite{nie03}) and their lobe orientation from \citet{tak08} in Table \ref{tab:table3}, and they are also plotted in figure \ref{fig:PAs}.  ``Outflow'' column in Table \ref{tab:table3} lists only outflows with measurable lobe directions.  
We wonder if magnetic field orientation is related to any of the above orientations.  \citet{poi10} have shown the distribution of the magnetic field orientation for the OMC-2/3 region.  They concluded that no correlation is evident between the relative orientation of jets or outflows and the magnetic field. The PA of the magnetic field is $\sim$ 135$\degree$ near MMS2 and MMS7, while it is $\sim$ 90$\degree$ near MMS9.  Due to the limited number of outflows having well-defined orientations, it is hard to reach any clear conclusion, but at least we can say that we do not see any very strong correlation among these orientations in Tables \ref{tab:table2} and \ref{tab:table3}.

\subsection{Dependence on the Employed Molecular Line}
Lastly, we discuss how our result could depend on selected molecular lines.
If we analyze the velocity gradient on the basis of observations with different molecular lines, the results could be different (e.g., \cite{goo93}).  We used N$_2$H$^+$ observations for Orion cores, while \citet{goo93} used NH$_3$ for cold dark cloud cores.  In general, N$_2$H$^+$ and NH$_3$ show rather similar distributions, compared with other molecules such as CCS, CS, C$^{18}$O, and C$^{17}$O (e.g., \cite{taf02,aik01} for L1544 and other starless cold dark cloud cores).  Therefore, we expect N$_2$H$^+$ and NH$_3$ to trace similar volumes.  On the other hand, \citet{hot04} reported changes in the abundance ratio of NH$_3$ to N$_2$H$^+$.  In the future, it is desirable to check our result by using the same molecular line, the same core identification, and velocity gradient measurement method with comparable linear spatial resolutions in pc.

\newpage

\section{Summary}

We have analyzed the specific angular momentum of molecular cloud cores in the Orion A giant molecular cloud using the N$_2$H$^+$ data of \citet{tat08}.  We have measured the velocity gradient using four position velocity diagrams passing through the core centers with PA = 0, 45, 90, and 135 degrees, and made a sinusoidal fitting on the velocity gradient-PA diagram.  By comparing the $J/M-R$ relation, 
we marginally found that 
the Orion N$_2$H$^+$ cores have systematically larger $J/M$ than cold dark cloud cores of \citet{goo93}.  The logarithm of the ratio of rotational to gravitational energy is 
log $\beta$ = $-2.3\pm0.7$ and $-1.9\pm0.7$ for Orion N$_2$H$^+$ cores and cold dark cloud cores, respectively.  
The large-scale rotation of the $\int$-shaped filament of the Orion A GMC does not likely govern the core rotation at smaller scales. 
During the analysis, we also found that
cores elongated perpendicular to the large-scale filament are rare. 

\bigskip

We wish to thank an anonymous referee for very valuable constructive comments that have significantly improved the presentation of the paper.
\begin{longtable}{lrrlrrlcccccccc}

  \caption{N$_2$H$^+$ cores in the Orion A GMC }\label{tab:catalog}
  \hline              
No.	&	R. 	&	A.	&	(J2000)	&		&	Dec.	&	(J2000)	&	$T_k$	&	$V_{LSR}$	&	$\Delta v$			&	$\Delta v_{TOT}$	&	$R$	&	$M$	&	$M_{vir}$	&	$\alpha_{vir}$	\\
\hline

	&	h	&	m	&	s	&	$\degree$	&	$\arcmin$	&	$\arcsec$	&	K	&	km s$^{-1}$	&	km s$^{-1}$			&	km s$^{-1}$	&	pc	&	$M_{\odot}$	&	$M_{\odot}$	&		\\
	
\endfirsthead
  \hline
	
\endhead
  \hline
\endfoot 
  \hline
\endlastfoot
  \hline
1	&	5	&	35	&	06.2	&	-4	&	54 	&	24 	&		&	11.1 	&	0.65 	$\pm$	0.05 	&		&	0.049 	&		&		&		\\
2	&	5	&	35	&	06.1	&	-4	&	56 	&	07 	&		&	11.2 	&	0.62 	$\pm$	0.03 	&		&	0.088 	&	22	&		&		\\
3	&	5	&	35	&	29.4	&	-4	&	58 	&	31 	&	30	&	12.3 	&	1.31 	$\pm$	0.10 	&	1.50 	&	0.083 	&	39	&	39	&	1.02	\\
4	&	5	&	35	&	19.8	&	-5	&	00 	&	53 	&	16	&	11.3 	&	0.57 	$\pm$	0.01 	&	0.79 	&	0.084 	&	43	&	11	&	0.25	\\
5	&	5	&	35	&	26.8	&	-5	&	01 	&	13 	&	22	&	11.5 	&	0.79 	$\pm$	0.05 	&	1.01 	&	0.059 	&	10	&	13	&	1.33	\\
6	&	5	&	35	&	25.4	&	-5	&	02 	&	36 	&	24	&	11.1 	&	0.51 	$\pm$	0.02 	&	0.83 	&	0.099 	&	47	&	15	&	0.31	\\
7	&	5	&	35	&	26.7	&	-5	&	05 	&	00 	&	28	&	11.6 	&	0.47 	$\pm$	0.01 	&	0.85 	&	0.091 	&	41	&	14	&	0.33	\\
8	&	5	&	35	&	32.3	&	-5	&	06 	&	02 	&	34	&	11.8 	&	0.74 	$\pm$	0.04 	&	1.07 	&	0.049 	&		&	12	&		\\
9	&	5	&	35	&	23.9	&	-5	&	07 	&	25 	&	26	&	11.9 	&	0.86 	$\pm$	0.04 	&	1.10 	&	0.098 	&	47	&	25	&	0.53	\\
10	&	5	&	35	&	26.7	&	-5	&	10 	&	09 	&	28	&	11.3 	&	1.23 	$\pm$	0.02 	&	1.42 	&	0.095 	&	57	&	40	&	0.7	\\
11	&	5	&	35	&	22.6	&	-5	&	10 	&	09 	&		&	11.7 	&	0.62 	$\pm$	0.02 	&		&	0.052 	&	9	&		&		\\
12	&	5	&	35	&	22.7	&	-5	&	12 	&	32 	&	28	&	11.0 	&	0.93 	$\pm$	0.02 	&	1.17 	&	0.104 	&	77	&	30	&	0.38	\\
13	&	5	&	35	&	21.2	&	-5	&	14 	&	36 	&	19	&	10.8 	&	0.62 	$\pm$	0.02 	&	0.86 	&	0.094 	&	39	&	14	&	0.37	\\
14	&	5	&	35	&	08.8	&	-5	&	18 	&	41 	&	24	&	9.0 	&	0.62 	$\pm$	0.03 	&	0.91 	&	0.083 	&	21	&	14	&	0.68	\\
15	&	5	&	35	&	15.8	&	-5	&	19 	&	26 	&	30	&	9.9 	&	1.31 	$\pm$	0.02 	&	1.50 	&	0.111 	&	117	&	53	&	0.45	\\
16	&	5	&	35	&	08.9	&	-5	&	20 	&	22 	&		&	8.6 	&	2.05 	$\pm$	0.09 	&		&	0.105 	&		&		&		\\
17	&	5	&	35	&	10.3	&	-5	&	21 	&	25 	&		&	8.1 	&	1.08 	$\pm$	0.05 	&		&	0.074 	&		&		&		\\
18	&	5	&	35	&	06.1	&	-5	&	22 	&	46 	&		&	7.4 	&	2.06 	$\pm$	0.14 	&		&	0.057 	&		&		&		\\
19	&	5	&	35	&	12.9	&	-5	&	24 	&	10 	&	61	&	6.6 	&	2.12 	$\pm$	0.07 	&	2.37 	&	0.072 	&		&	85	&		\\
20	&	5	&	35	&	04.7	&	-5	&	24 	&	13 	&	26	&	8.6 	&	1.92 	$\pm$	0.10 	&	2.03 	&	0.043 	&		&	38	&		\\
21	&	5	&	35	&	15.7	&	-5	&	25 	&	54 	&	55	&	8.2 	&	1.15 	$\pm$	0.06 	&	1.52 	&	0.062 	&	13	&	30	&	2.4	\\
22	&	5	&	35	&	14.3	&	-5	&	26 	&	56 	&	40	&	8.8 	&	1.72 	$\pm$	0.07 	&	1.92 	&	0.074 	&		&	57	&		\\
23	&	5	&	35	&	02.0	&	-5	&	36 	&	10 	&	24	&	7.3 	&	0.32 	$\pm$	0.01 	&	0.74 	&	0.115 	&	59	&	13	&	0.22	\\
24	&	5	&	35	&	04.8	&	-5	&	37 	&	32 	&	16	&	8.8 	&	0.72 	$\pm$	0.03 	&	0.90 	&	0.083 	&	41	&	14	&	0.34	\\
25	&	5	&	34	&	56.6	&	-5	&	41 	&	39 	&		&	3.7 	&	0.48 	$\pm$	0.03 	&		&	0.084 	&		&		&		\\
26	&	5	&	34	&	57.7	&	-5	&	43 	&	41 	&	24	&	7.1 	&	0.41 	$\pm$	0.03 	&	0.78 	&	0.062 	&		&	8	&		\\
27	&	5	&	34	&	56.3	&	-5	&	46 	&	05 	&	24	&	5.5 	&	1.11 	$\pm$	0.08 	&	1.29 	&	0.074 	&	20	&	26	&	1.27	\\
28	&	5	&	35	&	08.8	&	-5	&	51 	&	57 	&	16	&	7.0 	&	0.38 	$\pm$	0.05 	&	0.66 	&	0.081 	&		&	8	&		\\
29	&	5	&	35	&	00.7	&	-5	&	55 	&	40 	&		&	8.0 	&	0.49 	$\pm$	0.05 	&		&	0.062 	&		&		&		\\
30	&	5	&	35	&	09.0	&	-5	&	55 	&	41 	&	34	&	7.5 	&	0.64 	$\pm$	0.04 	&	1.01 	&	0.052 	&	6	&	11	&	1.77	\\
31	&	5	&	35	&	12.8	&	-5	&	58 	&	06 	&	16	&	7.6 	&	0.83 	$\pm$	0.09 	&	0.99 	&	0.083 	&		&	17	&		\\
32	&	5	&	35	&	28.1	&	-6	&	00 	&	09 	&		&	7.3 	&	0.35 	$\pm$	0.04 	&		&	0.157 	&	67	&		&		\\
33	&	5	&	36	&	12.4	&	-6	&	10 	&	44 	&	34	&	8.2 	&	0.62 	$\pm$	0.06 	&	1.00 	&	0.057 	&	11	&	12	&	1.05	\\
34	&	5	&	36	&	24.7	&	-6	&	14 	&	11 	&	21	&	8.2 	&	0.92 	$\pm$	0.16 	&		&	0.072 	&		&		&		\\
\end{longtable}


\begin{longtable}{lrrrrrrrrr}

  \caption{Velocity gradient, angular momentum of the Orion A GMC N$_2$H$^+$ cores}\label{tab:table2}
  \hline              
No.	&	V Grad	&		&		&		&	V Grad PA (fit)			&	V Grad (fit)			&	$v_{rot}$			&	J/M	&	$\beta$	\\
\hline

	&	PA = 0$\degree$	&	PA = 45$\degree$	&	PA = 90$\degree$	&	PA = 135$\degree$	&				&				&				&		&		\\
\hline

	&	\tiny{km s$^{-1}$ pc$^{-1}$}	&	\tiny{km s$^{-1}$ pc$^{-1}$}	&	\tiny{km s$^{-1}$ pc$^{-1}$}	&	\tiny{km s$^{-1}$ pc$^{-1}$}	&	$\degree$			&	km s$^{-1}$ pc$^{-1}$			&	km s$^{-1}$			&	cm$^2$ s$^{-1}$	&		\\

\endfirsthead
  \hline
	
\endhead
  \hline
\endfoot 
  \hline
\endlastfoot
  \hline
1	&	0	&	1.26	&	2.19	&	2.52	&	100.5	$\pm$	4.8	&	2.5	$\pm$	0.2	&	0.12 	$\pm$	0.01 	&	7.4E+20	&		\\
2	&	-0.69	&	0	&	0	&	3.05	&	143.8	$\pm$	17.1	&	1.8	$\pm$	0.5	&	0.16 	$\pm$	0.05 	&	1.7E+21	&	7.6E-03	\\
3	&	-0.96	&	1.26	&	1.1	&	-0.19	&	87.9	$\pm$	25.5	&	0.9	$\pm$	0.4	&	0.07 	$\pm$	0.03 	&	7.6E+20	&	9.3E-04	\\
4	&	0	&	-0.78	&	-1.1	&	0.44	&	237.3	$\pm$	17.9	&	0.8	$\pm$	0.2	&	0.07 	$\pm$	0.02 	&	7.2E+20	&	7.3E-04	\\
5	&		&		&		&		&				&				&				&		&		\\
6	&	0.27	&	0	&	-0.96	&	0	&	286	$\pm$	23.4	&	0.5	$\pm$	0.2	&	0.05 	$\pm$	0.02 	&	6.0E+20	&	4.0E-04	\\
7	&	-0.96	&	-1.16	&	-0.41	&	0.68	&	198.4	$\pm$	3.4	&	1.2	$\pm$	0.1	&	0.10 	$\pm$	0.01 	&	1.2E+21	&	1.9E-03	\\
8	&	6.44	&	0	&	-2.06	&	-3.88	&	332.4	$\pm$	8.5	&	5.2	$\pm$	0.7	&	0.26 	$\pm$	0.04 	&	1.6E+21	&		\\
9	&	-5.07	&	-7.85	&	-5.89	&	0.68	&	224.6	$\pm$	1.9	&	7.8	$\pm$	0.2	&	0.77 	$\pm$	0.02 	&	9.3E+21	&	9.5E-02	\\
10	&	1.64	&	1.74	&	-1.64	&	-3.2	&	332.5	$\pm$	7.8	&	2.9	$\pm$	0.4	&	0.27 	$\pm$	0.04 	&	3.2E+21	&	9.7E-03	\\
11	&	1.23	&	0.63	&	0	&	0	&	14.9	$\pm$	12.2	&	0.9	$\pm$	0.2	&	0.05 	$\pm$	0.01 	&	3.0E+20	&	9.7E-04	\\
12	&	2.33	&	0	&	-1.78	&	-0.53	&	321.4	$\pm$	16.3	&	1.7	$\pm$	0.5	&	0.18 	$\pm$	0.05 	&	2.3E+21	&	3.3E-03	\\
13	&	-0.69	&	-0.78	&	-2.47	&	0	&	247.8	$\pm$	13.8	&	1.6	$\pm$	0.4	&	0.15 	$\pm$	0.04 	&	1.8E+21	&	4.5E-03	\\
14	&	1.92	&	2.81	&	0	&	-1.55	&	10.1	$\pm$	6.8	&	2.5	$\pm$	0.3	&	0.21 	$\pm$	0.03 	&	2.2E+21	&	1.3E-02	\\
15	&	-0.96	&	-0.29	&	-1.78	&	-1.02	&	260.7	$\pm$	14.5	&	1.4	$\pm$	0.3	&	0.16 	$\pm$	0.04 	&	2.1E+21	&	1.8E-03	\\
16	&	0.96	&	-3.97	&	0	&	1.16	&	216.7	$\pm$	14.5	&	1.4	$\pm$	0.3	&	0.15 	$\pm$	0.03 	&	1.9E+21	&		\\
17	&	2.33	&	0	&	0	&	-4.75	&	329.4	$\pm$	12.4	&	3.3	$\pm$	0.7	&	0.24 	$\pm$	0.05 	&	2.2E+21	&		\\
18	&		&		&		&		&				&				&				&		&		\\
19	&		&		&		&		&				&				&				&		&		\\
20	&	1.03	&	-7.37	&	0	&	3.39	&	203.1	$\pm$	29.7	&	3.6	$\pm$	1.9	&	0.16 	$\pm$	0.08 	&	8.4E+20	&		\\
21	&	0	&	-6.4	&	0	&	-1.74	&	240.3	$\pm$	23.4	&	3.3	$\pm$	1.3	&	0.20 	$\pm$	0.08 	&	1.5E+21	&	1.6E-02	\\
22	&		&		&		&		&				&				&				&		&		\\
23	&	1.1	&	1.45	&	0	&	0	&	25.8	$\pm$	10.2	&	1.2	$\pm$	0.2	&	0.13 	$\pm$	0.03 	&	1.9E+21	&	2.6E-03	\\
24	&	-4.93	&	-2.23	&	0.27	&	5.14	&	167.1	$\pm$	4.1	&	5.2	$\pm$	0.3	&	0.43 	$\pm$	0.03 	&	4.4E+21	&	2.9E-02	\\
25	&		&		&		&		&				&				&				&		&		\\
26	&	0.96	&	-0.78	&	-1.37	&	-1.55	&	296.6	$\pm$	3.4	&	1.6	$\pm$	0.1	&	0.10 	$\pm$	0.01 	&	7.7E+20	&		\\
27	&		&		&		&		&				&				&				&		&		\\
28	&	-0.69	&	0	&	0.62	&	1.94	&	136	$\pm$	8.3	&	1.4	$\pm$	0.2	&	0.11 	$\pm$	0.02 	&	1.1E+21	&		\\
29	&	2.47	&	1.65	&	2.47	&	0	&	45	$\pm$	8.4	&	2.5	$\pm$	0.4	&	0.16 	$\pm$	0.03 	&	1.2E+21	&		\\
30	&	0.96	&	4.46	&	1.58	&	1.45	&	61.8	$\pm$	10.2	&	3.3	$\pm$	0.6	&	0.17 	$\pm$	0.03 	&	1.1E+21	&	1.9E-02	\\
31	&	-0.82	&	-3.78	&	-2.88	&	-1.55	&	250.2	$\pm$	3.9	&	3.5	$\pm$	0.2	&	0.29 	$\pm$	0.02 	&	3.0E+21	&		\\
32	&	-0.96	&	0	&	0	&	1.45	&	152.7	$\pm$	10.8	&	1.2	$\pm$	0.2	&	0.18 	$\pm$	0.04 	&	3.5E+21	&	6.0E-03	\\
33	&	-0.96	&	0	&	-3.7	&	-1.07	&	267.4	$\pm$	17.9	&	2.2	$\pm$	0.7	&	0.13 	$\pm$	0.04 	&	8.9E+20	&	6.3E-03	\\
34	&		&		&		&		&				&				&				&		&		\\
\end{longtable}


\begin{longtable}{lcclc}

  \caption{Position angle of the core elongation and the associated molecular outflow for Orion A GMC N$_2$H$^+$ cores}\label{tab:table3}
  \hline              

No.	&	Elongation PA			&	Angle between V Grad and Elongation	&	Outflow	&	Outflow PA	\\
  \hline

	&	$\degree$			&	$\degree$	&		&	$\degree$	\\
 
\endfirsthead
  \hline
	
\endhead
  \hline
\endfoot 
  \hline
\endlastfoot
  \hline
1	&				&		&		&		\\
2	&	155.2 	$\pm$	4.3 	&	11.4 	&		&		\\
3	&	170.9 	$\pm$	7.1 	&	83.0 	&	OMC-3 SIMBA~a, SIMBA~c	&	50	\\
4	&	131.2 	$\pm$	6.7 	&	73.9 	&	OMC-3 MMS2	&	90	\\
5	&				&		&		&		\\
6	&	153.7 	$\pm$	4.5 	&	47.7 	&	OMC-3 MMS7	&	90	\\
7	&				&		&	OMC-3 MMS9	&	80	\\
8	&	123.7 	$\pm$	5.9 	&	28.7 	&		&		\\
9	&	164.6 	$\pm$	0.8 	&	60.0 	&	OMC-2 FIR1b	&	160	\\
10	&				&		&	OMC-2 FIR3	&	30	\\
11	&	36.1 	$\pm$	16.9 	&	21.2 	&		&		\\
12	&	34.7 	$\pm$	4.0 	&	73.3 	&	OMC-2 FIR6b	&	60	\\
13	&	36.7 	$\pm$	4.2 	&	31.1 	&		&		\\
14	&				&		&		&		\\
15	&	6.0 	$\pm$	4.2 	&	74.7 	&		&		\\
16	&	172.5 	$\pm$	3.4 	&	44.2 	&		&		\\
17	&	153.8 	$\pm$	0.6 	&	4.4 	&		&		\\
18	&				&		&		&		\\
19	&				&		&	Orion-S	&	35	\\
20	&	9.4 	$\pm$	18.4 	&	13.7 	&		&		\\
21	&				&		&		&		\\
22	&				&		&		&		\\
23	&	17.3 	$\pm$	1.8 	&	8.5 	&		&		\\
24	&	41.6 	$\pm$	7.5 	&	54.5 	&		&		\\
25	&				&		&		&		\\
26	&	149.1 	$\pm$	4.1 	&	32.5 	&		&		\\
27	&				&		&		&		\\
28	&	159.8 	$\pm$	3.7 	&	23.8 	&		&		\\
29	&				&		&		&		\\
30	&				&		&		&		\\
31	&				&		&		&		\\
32	&	18.2 	$\pm$	13.2 	&	45.5 	&		&		\\
33	&	130.3 	$\pm$	5.4 	&	42.9 	&		&		\\
34	&				&		&		&		\\
\end{longtable}





\begin{thebibliography}{}
\bibitem[Aikawa et al.(2001)]{aik01} Aikawa Y., 
Ohashi, N., Inutsuka, S.,
Herbst, E., \& Takakuwa, S. 2001, \apj, 552, 639
\bibitem[Aso et al.(2000)]{aso00} Aso, Y., Tatematsu, K., 
Sekimoto, Y., 
Nakano, T., Umemoto, T., Koyama, K., \& Yamamoto, S. 
2000, \apjs, 131, 465
\bibitem[Ballesteros-Paredes et al.(2011)]{bal11} Ballesteros-Paredes, J., Hartmann, L.~W., V\'{a}zquez-Semadeni, E., Heitsch, F., \& Zamora-Avil\'{e}s, M.~A. 2011 \mnras, 411, 65
\bibitem[Bergin et al.(2001)]{ber01} Bergin, E.A., Ciardi, D.R., 
Lada, C.J.,
Alves, J., 
\& Lada, E.A. 2001, \apj, 557, 209
\bibitem[Bergin et al.(2002)]{ber02} Bergin, E.A., Alves, J., 
Huard, T.L.,
\& Tafalla, M. 2002, \apj, 570, L101
\bibitem[Bergin \& Tafalla(2007)]{ber07} Bergin, E.~A.,
\& Tafalla, M. 2007, \araa, 45, 339
\bibitem[Blitz(1990)]{bli90} 
Blitz, L. 1990, ASPC, 12, 273
\bibitem[Blitz(1993)]{bli93} 
Blitz, L. 1993, in Protostars and Planets III, 
ed.\ E.~H. Levy \& J.~I. Lunine (Tucson: Univ Ariz. Press), 125
\bibitem[Blitz \& Rosolowsky(2006)]{bli06} 
Blitz, L., \& Rosolowsky, E. 2006, 
\apj, 650, 933
\bibitem[Bodenheimer(1995)]{bod95} 
Bodenheimer, P. 1995, \araa, 33, 199
\bibitem[Caselli, Myers, \& Thaddeus(1995)]{cas95a} Caselli, P., Myers, P.C.,
\& Thaddeus, P. 1995, \apj, 455, L77
\bibitem[Caselli \& Myers (1995)]{cas95b} Caselli, P., \& Myers, P.~C.
 1995,\apj, 446, 665
\bibitem[Chi\`{e}ze(1987)]{chi87} Chi\`{e}ze, J.~P.
1987, \aap, 171, 225
\bibitem[Chini et al.(1997)]{chi97} Chini, R., et al.
1997, \apj, 474, L135
\bibitem[Crutcher(2012)]{cru12} Crutcher, R.M.
2012, \araa, 50, 29
\bibitem[Danby et al.(1988)]{dan88} Danby, G., et al. 1988, 
\mnras, 235, 229
\bibitem[Field(1978)]{fie78} 
Field, G.~B. 1978, 
in Protostars and Planets, ed.\ T. Gehrels (Tucson: Univ. Ariz. Press), 243
\bibitem[Fuller \& Myers(1992)]{ful92} 
Fuller, G.~A., \& Myers, P.~C.
1992, \apj, 384, 523
\bibitem[Genzel \& Stutzki(1989)]{gen89} Genzel, R., 
\& Stutzki, R. 1989, \araa, 27, 41
\bibitem[Goldsmith \& Arquilla (1985)]{gol85}
Goldsmith, P.F., \& Arquilla, R.A. 1985, in Protostars and Planets II, ed.\ D.~C. Black, M.~S. Mathews (Tucson: Univ. Ariz. Press), 137
\bibitem[Goodman et al.(1993)]{goo93} 
Goodman, A.~A., Benson, P.~J., Fuller, G.~A., \& Myers, P.~C.
1993, \apj, 406, 528
\bibitem[Heyer et al.(2009)]{hey09} Heyer, M.~H, Krawczyk, C., Duval, J., \& Jackson, J.~M. 2009, \apj, 699, 1092
\bibitem[Hotzel, Harju, \& Walmsley(2004)]{hot04} Hotzel, S., Harju, J., \& Walmsley, C.~M. 2004, \aap, 415, 1065
\bibitem[Ikeda et al.(2007)]{ike07} Ikeda, N., Sunada, K.,  
\& Kitamura, Y. 2007, \apj, 665, 1194
\bibitem[Imara \& Blitz(2011)]{ima11a} Imara, N.,  
\& Blitz, L. 2011, \apj, 732, 78
\bibitem[Imara et al.(2011)]{ima11b} Imara, N., Bigiel, F., 
\& Blitz, L. 2011, \apj, 732, 79
\bibitem[Inoue \& Inutsuka(2012)]{ino12} Inoue, T., \& Inutsuka, S. 2012,
\apj, 759, 35  
\bibitem[Kim et al.(2008)]{kim08} Kim, M.~K., et al. 
2008, PASJ, 60, 991
\bibitem[Kutner et al.(1977)]{kut77} Kutner, M.~L., Tucker, K.~D., Chin, G., \& Thadeus, P. 1977, \apj, 215, 521 
2008, PASJ, 60, 991
\bibitem[Larson(1981)]{lar81} Larson, R.~B. 1981, \mnras, 194, 809
\bibitem[Lee, Bergin, \& Evans(2004)]{lee04} Lee, J-.E., Bergin, E.A.,
\& Evans, N.J., II  2004, \apj, 617, 360
\bibitem[MacLaren et al.(1988)]{mac88} MacLaren, I, Richardson, K.~M., \&
Wolfendale, A.~W. 1988, \apj, 333, 821 
\bibitem[Matsumoto \& Hanawa(2003)]{mat03} 
Matsumoto, T. \& Hanawa, T. 2003, \apj, 595, 913 
\bibitem[Matsumoto \& Tomisaka(2004)]{mat04} 
Matsumoto, T. \& Tomisaka, K. 2004, \apj, 616, 266 
\bibitem[Miyama, Hayashi,  \& Narita(1984)]{miy84} 
Miyama, S.~M., Hayashi, C., \& Narita, S. 1984, \apj, 279, 621 
\bibitem[Myers (1978)]{mye78} Myers, P.C. 1978,
\apj, 225, 380
\bibitem[Myers (1983)]{mye83} Myers, P.C. 1983,
\apj, 270, 105
\bibitem[Nielbock et al.(2003)]{nie03}
Nielbock, M., Chini, R., \& Muller, S.~A.~H. 2003, \aap, 408, 245 
\bibitem[Onishi et al.(2002)]{oni02} Onishi, T., Mizuno, A., Kawamura, A., Tachihara, K., \& Fukui, Y. 2002,
\apj, 575, 950
\bibitem[Poidevin et al.(2010)]{poi10} Poidevin, F., Bastien, P., \&
Matthews, B.~C. 2010, \apj, 716, 893
\bibitem[Schmidt-Burgk et al.(1990)]{sch90} Schmid-Burgk, J., G\"{u}sten, R., Mauersberger, R., Schulz, A., \& Wilson, T.~L. 1990, \apj, 362, L25
\bibitem[Shimajiri et al.(2008)]{shi08} Shimajiri, Y., Takahashi, S., 
Takakuwa, S., Saito, M., \& Kawabe, R. 2008, \apj,
638, 255
\bibitem[Shu et al.(1987)]{shu87} 
Shu, F.~H., Adams, F.~C., \& Lizano, S. 1987,
\araa, 25, 23
\bibitem[Tafalla et al.(2002)]{taf02} Tafalla, M., Myers, P.~C.,
Caselli, P., Walmsley, C.~M., \& Comito, C., 2002
\apj, 569, 815
\bibitem[Takahashi et al.(2008)]{tak08} Takahashi, S.,
Saito, M., Ohashi, N., Kusakabe, N., Takakuwa, S., 
Shimajiri, Y., Tamura, M., \& Kawabe, R. 2008, 
\apj, 688, 344
\bibitem[Tatematsu et al.(1993)]{tat93} Tatematsu, K., et al. 1993, 
\apj, 404, 643
\bibitem[Tatematsu (1999)]{tat99} Tatematsu, K. 1999, in Star Formation 1999, ed.\ T. Nakamoto (Nobeyama: Nobeyama Radio Observatory), 72
\bibitem[Tatematsu et al.(2008)]{tat08} Tatematsu, K., Kandori, R.,
Umemoto, T., \& Sekimoto, Y. 2008, 
PASJ, 60, 407
\bibitem[Tatematsu et al.(2014a)]{tat14} Tatematsu, K., Ohashi, S., 
Umemoto, T., Lee, J.-E., Hirota, T., Yamamoto, S., et al. 2014, 
\pasj, 66, 16
\bibitem[Traficante et al.(2016)]{tra16} Traficante, A., Fuller, G.~A., Smith, R., Billot, N.,  Duarte-Cabral, A. Peretto, N., et al. \mnras, in press, arXiv:1511.03670
\bibitem[Tsuboi \& Miyazaki(2012)]{tsu12} Tsuboi, M., \& Miyazaki, A. 2012,
\pasj, 64, 111
\bibitem[Tsuribe \& Inutsuka(1999)]{tsu99} Tsuribe, T., \& Inutsuka, S. 1999,
\apj, 523, L155 
\bibitem[Turner (1988)]{tur88} Turner, B.~E. 1988, in Galactic and 
Extragalactic Radio Astronomy, ed.\ G.~L. Vershuur \& K.~I. Kellermann 
(2d ed.; New York: Springer-Verlag), 154
\bibitem[Wilson et al.(1999)]{wil99} Wilson, T.~L., Mauersberger, R., 
Gensheimer, P.~D., Muders, D.,
\& Bieging, J.~H. 1999, \apj, 525, 343
\bibitem[Womack, Ziurys, \& Sage(1993)]{wom93} Womack, M., Ziurys, L.M., 
\& Sage, L.J. 1993, \apj, 406, L29
\bibitem[Yu et al.(2000)]{yu00} Yu, K.~C., Billawala, Y., Smith, M.~D., 
Bally, J. \& Butner, H.~M. 2000, \apj, 120, 1974
\end{thebibliography}
\end{document}